\documentclass[12pt]{article}
\usepackage{graphicx}
\usepackage{url} 
\newcommand{\blind}{0}

\usepackage{amsmath, amsthm, amssymb,bm}  
\usepackage{graphicx, subcaption, caption, float, arydshln}  
\usepackage{algorithm}
\usepackage{algorithmicx}
\usepackage{algpseudocode}
\usepackage{natbib}
\usepackage{setspace}

\usepackage{xcolor}

\newcommand{\Y}{\bm Y}

\newcommand{\x}{\bm{x}}

\newcommand{\R}{\bm{R}}

\DeclareMathOperator*{\argmax}{arg\,max}

\usepackage{xr}
\makeatletter
\newcommand*{\addFileDependency}[1]{
\typeout{(#1)}
\@addtofilelist{#1}
\IfFileExists{#1}{}{\typeout{No file #1.}}
}\makeatother
\newcommand*{\myexternaldocument}[1]{%
\externaldocument{#1}%
\addFileDependency{#1.tex}%
\addFileDependency{#1.aux}%
}
\myexternaldocument{supp}

\begin{document}

\def\spacingset#1{\renewcommand{\baselinestretch}%
{#1}\small\normalsize} \spacingset{1}


\if0\blind
{
  \title{\bf Active Learning via\\ Heteroskedastic Rational Kriging}
  \author{Shangkun Wang and V. Roshan Joseph\thanks{Corresponding author: roshan@gatech.edu}\\
    H. Milton Stewart School of Industrial and Systems Engineering,\\ Georgia Institute of Technology, Atlanta, GA 30332\\
}
  \maketitle
} \fi

\if1\blind
{
  \bigskip
  \bigskip
  \bigskip
  \begin{center}
    {\LARGE\bf Active learning with HRK}
\end{center}
  \medskip
} \fi

\bigskip
\begin{abstract}
Active learning methods for emulating complex computer models that rely on stationary Gaussian processes tend to produce design points that uniformly fill the entire experimental region, which can be wasteful for functions which vary only in small regions. In this article, we propose a new Gaussian process model that captures the heteroskedasticity of the function. Active learning using this new model can place design points in the more interesting regions of the response surface, and thus obtain surrogate models with better accuracy. The proposed active learning method is compared with the state-of-the-art methods using simulations and two real datasets. It is found to have comparable or better performance relative to other non-stationary Gaussian process-based methods, but faster by orders of magnitude.

\end{abstract}

\noindent%
{\it Keywords:} Computer experiments, Emulation, Gaussian process, Non-stationarity, Sequential design, Surrogate model. 


\spacingset{1.5} 
\section{Introduction}
\label{sec:intro}
Computer models are widely used in science and engineering to simulate physical systems. For example, \cite{mak2018efficient} used large eddy simulations to understand the liquid oxygen flow inside a rocket injector, \cite{krishna2024adaptive} used density functional theory codes to map the potential energy surface of crystal structures, and \cite{huang2021bayesian} used reaction-diffusion models to study the properties of hybrid polymer membranes produced from a vapor phase infiltration process. These computer models are expensive to evaluate, and therefore, they are replaced with easy-to-evaluate surrogate models. Surrogate models are created by running computer models using an experimental design and then fitting some statistical models such as kriging or Gaussian process (GP) to the data \citep{santner2019design, gramacy2020surrogates}.

Space-filling designs \citep{joseph2016space} are widely used as experimental designs for computer experiments. These designs possess some robustness properties and work well over a wide class of computer models. However, it may not be the best experimental design to use for a given computer model. For example, if the computer model is sensitive to the inputs only in a small region of the experimental space and constant everywhere else, then it is wasteful to use a space-filling design to evenly cover the entire experimental region. Better experimental designs for a given computer model can be created using sequential design strategies, also known as active learning. In an active learning strategy, we will start with a small space-filling design, approximately learn about the response surface, and then add points sequentially in regions where the approximation errors are likely to be large.

A crucial step in an active learning strategy is to identify the region where the current surrogate model gives a poor approximation to the true response surface. This is not easy because we do not know the true response surface. This is where the kriging/GP model comes in handy. These models can not only provide a prediction but can also probabilistically quantify the approximation errors. Therefore, their uncertainty estimates can be used to find regions that are likely to have high prediction errors. Although kriging/GP models are widely used for active learning \citep[Sec. 6.2]{gramacy2020surrogates}, they have a major drawback.

\begin{figure}
    \centering
    \begin{subfigure}[b]{0.35\textwidth}
        \centering
        \includegraphics[width=1\textwidth]{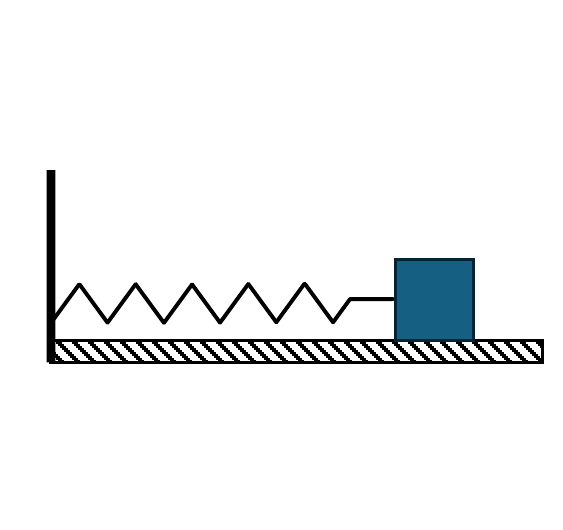}
    \caption{}
        \label{fig:uos}
    \end{subfigure}
    \begin{subfigure}[b]{0.35\textwidth}  
        \centering 
        \includegraphics[width=\textwidth]{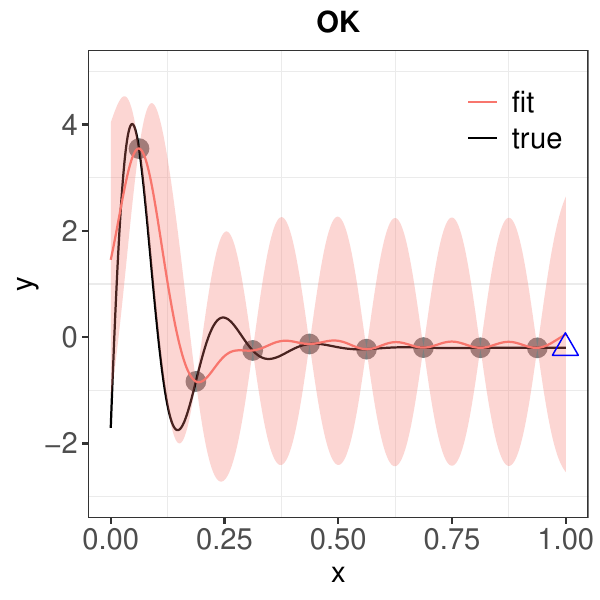}
        \caption{}    
        \label{fig:ExpTrig_OK_seq}
    \end{subfigure}
    \caption{(a) Underdamped oscillator system and its response over time; (b) Ordinary kriging fitted on eight design points denoted as gray. The red solid lines denote the predicted mean and the shaded region is the 95\% confidence intervals. Black line is the true response. The blue triangle is the next design point selected by ALM. }
\end{figure}

To understand the limitation of GP models, consider an underdamped oscillator system shown in Figure \ref{fig:uos}. In this system, the moving part oscillates on a rough surface, with its amplitude decreasing exponentially until the system comes to rest. Suppose eight points are sampled using a minimax design, and the data are obtained (shown as gray points in the figure). A GP model (also known as ordinary kriging) can be fitted to this data as follows. Let $y=f(\bm x)$, $\bm x\in \mathbb{R}^p$ be the true function. Assume a GP prior for the function $f(\bm x)\sim GP(\mu,\nu^2 R(\cdot))$, where $\mu$ is the mean, $\nu^2$ the variance, and $R(\cdot)$ the correlation function defined as $cor\{f(\bm x_i),f(\bm x_j)\}=R(\bm x_i-\bm x_j)$. Now, given the design $\{\bm x_1,\ldots,\bm x_n\}$ and the data $\Y_n =[f(\x_1),\ldots,f(\x_n)]'$, the posterior distribution of $f(\bm x)$ for a new $\bm x$ is given by
\begin{align}
    f(\x)|\Y_n &\sim \mathcal{N}\left(\hat{y}(\x), s^2(\x)\right),\\
    \hat{y}(\x) &= \hat{\mu}+\bm{r}(\x)'\R^{-1}(\Y_n-\hat{\mu}\bm{1}_n), \label{eq:OK}\\
    s^2(\x) &= \nu^2\left\{1-\bm{r}(\x)'\R^{-1}\bm{r}(\x)\right\},
\end{align}
where $\bm{R}=\{R(\x_i-\x_j)\}_{n\times n}$ is the correlation matrix, $\bm{r}(\x) = (R(\x-\x_1),\dots,R(\x-\x_n))'$,  $\bm{1}_n$ is a vector of $1$'s, and $\hat{\mu}= \bm{1}_n'\R^{-1}\Y_n/\bm{1}_n'\R^{-1}\bm{1}_n$ is the generalized least squares estimate of $\mu$. Suppose we use a Gaussian correlation function 
\begin{equation*}
    R(\bm u,\bm v) = \exp{\left(-\sum_{l=1}^p\frac{(u_l-v_l)^2}{\theta_l^2}\right)},
\end{equation*}
where $\bm{\theta}=(\theta_1,\dots,\theta_p)'$ are the lengthscale parameters.
Figure \ref{fig:ExpTrig_OK_seq} shows the predictor $\hat{y}(\bm x)$ (solid red line) and the 95\% credible intervals $\hat{y}(\bm x)\pm 2s(\bm x)$ (shaded region), where the lengthscale parameter is estimated using empirical Bayes methods.

For active learning, it makes sense to sample the next point where the uncertainty is the largest, that is
\begin{equation}\label{eq:ALM}
    \bm x_{n+1}=\argmax_{\bm x\in \mathcal{X}} s^2(\bm x),
\end{equation}
where $\mathcal{X}\subset \mathbb{R}^p$ is the experimental region. As shown in \cite{mackay1992information}, this choice maximizes the entropy of the augmented design $\{\bm x_1,\ldots,\bm x_n, \bm x_{n+1}\}$. Therefore, in the active learning literature, this strategy is known as Active Learning MacKay (ALM). The ninth point obtained using this strategy is shown in Figure \ref{fig:ExpTrig_OK_seq}. However, the placement of this point seems counterintuitive. The true function fluctuates more on the left side and is almost flat on the right side. Thus, from the function approximation point of view, we should sample more points on the left side rather than the right side. This is a deficiency of the GP model caused by its stationarity assumption, that is, the function has constant variance throughout the space. This makes the credible intervals roughly the same size, whether it is computed on the left side or on the right side of the function. A closer look at the expression of $s(\bm x)$ also reveals the same issue, which is independent of $\bm Y_n$ and depends only on how the $\bm x_i$'s are distributed. \cite{Johnson1990} shows that a maximum entropy design is asymptotically equivalent to a maximin design. In other words, the ALM strategy will lead to a maximin-type design, which in fact could be worse than the optimal maximin design of the same size due to its one-point-at-a-time myopic construction. Thus, although our aim was to construct a better experimental design using active learning, we ended up in a design that could be much worse than a fixed space-filling design! To be fair, the active learning strategy does have some advantages over a fixed design: (i) we do not need to specify the size of the design beforehand, and we can stop adding points when the approximation is good enough, and (ii) since the lengthscale parameters are learned at each step, the design points will tend to a weighted maximin design which helps to fill the space of important variables. Nevertheless, these advantages are much less than what we had hoped to achieve from active learning.

We can overcome the aforementioned deficiency of the GP model by relaxing the stationarity assumption of variance. \cite{banerjee2003hierarchical} proposed to use the model $f(\bm x)=\mu+\nu(\bm x)Z(\bm x)$, where $Z(\bm x)\sim GP(0,R(\cdot))$ and $\nu^2(\bm x)$ is a deterministic function that captures heteroskedastic variance. However, it is not easy to postulate a deterministic model for the variance a priori \citep{ba2012composite}. Therefore, \cite{huang2011class} proposed to use another GP to model the variance: $\nu^2(\bm x)=\exp\{\alpha(\bm x)\}$, where $\alpha(\bm x)$ is a GP and the exponential is taken to ensure that the variance will be nonnegative. Extensions of this approach can be found in \cite{tolvanen2014} and \cite{heinonen2016non}. However, the product of two GPs make the computations challenging. Another approach to model heteroskedasticity is through divide and conquer, where local Gaussian processes are fitted to capture the local variability of the function. Two prominent examples are Treed Gaussian process \citep{gramacy2008bayesian} and local approximate Gaussian process \citep{gramacy2015local}. However, the localization introduces discontinuities in the prediction and may not work well unless we have large amounts of data. A yet another approach to introduce non-stationarity is through warped Gaussian process where the input space is transformed, enabling the function to be modeled effectively by a GP in the ``warped'' space \citep{sampson1992nonparametric, schmidt2003bayesian}. This idea is revived with the deep Gaussian process (DeepGP), which essentially stacks layers of latent GPs together in order to capture complicated 
non-stationary correlations \citep{damianou2013deep, dunlop2018deep}. \cite{sauer2023active} used DeepGP for active learning in computer experiments. Another idea to capture nonstationarity is dimension expansion with latent variables \citep{bornn2012modeling}, which was used for active learning in computer experiments by \cite{montagna2016computer}. However, these methods use Markov chain Monte Carlo (MCMC) for sampling-based inference, making the prediction and active learning slow, which limits their applicability. Please see \cite{sauer2023non} for a  more comprehensive review of non-stationary GPs. 

The main aim of this article is to develop a computationally efficient heteroskedastic GP model and use it to develop an active learning procedure that is fast and easy to use in computer experiments. We extend the recently proposed rational kriging \citep{joseph2024rational} for this purpose. Rational kriging is computationally efficient and contains enough degrees of freedom to seamlessly integrate heteroskedasticity into the modeling.

The article is organized as follows. Section \ref{sec:review} reviews rational kriging. In Section \ref{sec:hrk}, we propose heteroskedastic rational kriging and develop a computationally efficient procedure for its estimation. Its application to active learning is discussed in Section \ref{sec:active learning}. Section \ref{sec:simulation} contains extensive simulations to demonstrate the advantages of the proposed procedure compared to the state-of-the-art active learning strategies. Two real case studies are provided in Section \ref{sec:case}. We conclude with some remarks in Section \ref{sec:remarks}.

\section{Review of Rational Kriging}
\label{sec:review}
\cite{joseph2024rational} recently proposed a (generalized) rational predictor of the form
\[\hat{y}(\bm x)= \mu+\frac{\sum_{i=1}^n b_iR(\bm x-\bm x_i)}{c_0+\sum_{i=1}^n c_iR(\bm x-\bm x_i)},\]
where $\sum_{i=0}^n c_i^2=1$ and $c_i\ge 0$ for all $i=1,\ldots,n$. If $c_0=1$, then this reduces to the ordinary kriging (OK) predictor in (\ref{eq:OK}). The rational form of the predictor was motivated by previous work in kriging/GP \citep{joseph2006limit, kang2016kernel} and rational radial basis functions \citep{jakobsson2009rational, sarra2018rational, buhmann2020analysis}. This predictor can be obtained as the posterior mean of a GP of the form:
\begin{equation}\label{eq:RKmodel}
    y(\x) = \mu +\tau(\bm x) Z(\x),\; Z(\x)\sim GP(0,R(\cdot)),
\end{equation}
where
\begin{equation}\label{eq:tau}
    \tau(\bm x)=\frac{\nu}{c_0+\bm{r}(\x)'\bm{c}}.
\end{equation}
Let $\Tilde{\bm{c}}=(c_0,\; \bm{c}')'$ and $\Tilde{\R}=[\bm{1}_n,\R]$. Then, the posterior distribution of $y(\x)$ given the observations $\Y_n$ can be obtained as
\begin{equation}
    y(\x)|\Y_n, \Tilde{\bm{c}}, \nu, \mu, \theta \sim N \left(\hat{y}(\x), s^2(\x)\right),
\end{equation}
where
\begin{equation}\label{eq:grk_mean}
    \hat{y}(\x) = \mu + \frac{r(\x)'}{c_0+\bm{r}(\x)'\bm{c}}\R^{-1}\text{diag}(\Tilde{\R}\Tilde{\bm{c}})(\Y_n-\mu\bm{1}_n), 
\end{equation}
and
\begin{equation}\label{eq:grk_var}
    s^2(\x) = \frac{\nu^2}{\{c_0+\bm{r}(\x)'\bm{c}\}^2}\left\{{1-\bm{r}(\x)'\R^{-1}\bm{r}(\x)}\right\}.
\end{equation}

If we use a non-informative prior for $\mu$: $p(\mu)\propto 1$, then the posterior distribution of $\mu$ is given by
\begin{equation}\label{eq:mu_rk}
    \mu|\Y_n, \tilde{\bm{c}}, \nu,  \theta\sim \mathcal{N}\left(\frac{\tilde{\bm c}'\tilde{\bm R}'\bm R^{-1}\text{diag}(\tilde{\bm R}\tilde{\bm c})\bm Y_n}{\tilde{\bm c}'\tilde{\bm R}'\bm R^{-1} \tilde{\bm R}\tilde{\bm c}},\frac{\nu^2}{\tilde{\bm c}' \tilde{\bm R}'\bm R^{-1} \tilde{\bm R}\tilde{\bm c}}\right).
\end{equation}
The focus of \cite{joseph2024rational} was to improve the estimation of $\mu$. We can see that the posterior variance of $\mu$ can be minimized by maximizing the quadratic form $\tilde{\bm c}' \tilde{\bm R}'\bm R^{-1} \tilde{\bm R}\tilde{\bm c}$ with respect to $\tilde{\bm c}$ subject to the constraints $\tilde{\bm c}'\tilde{\bm c}=1$ and $\tilde{\bm c}\ge 0$. This optimization problem has an explicit solution given by the Perron eigenvector of $\Tilde{\R}'\R^{-1}\Tilde{\R}$.

\begin{figure}
    \centering
    \includegraphics[width=1\textwidth]{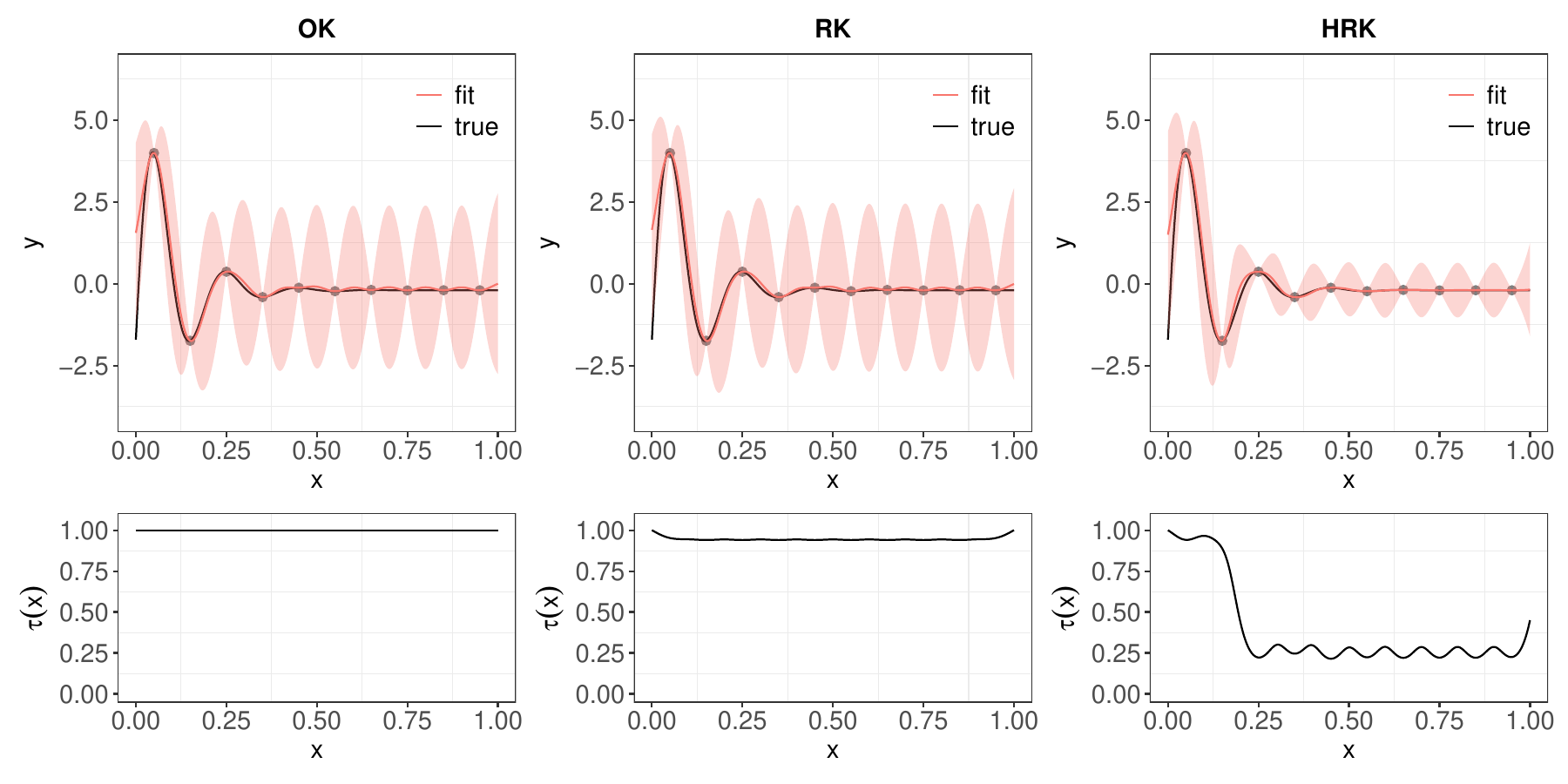}
    \caption{Upper panel: OK, RK and HRK on the underdamped oscillation system. The design is 10 equally spaced points shown in red. The red solid lines denote the predicted mean and the red shaded bands are the 95\% confidence intervals. Black line is the true response. Lower panel: the relative variance $\tau(x)$ over the input domain. It is scaled so that the maximum value is 1.}
    \label{fig:ExpTrig_fit}
\end{figure}

The middle panels of Figure \ref{fig:ExpTrig_fit} shows the rational kriging fit and $\tau(\bm x)$ in (\ref{eq:tau}) for the underdamped oscillation system example discussed in the introduction. We can see that the behavior of RK is almost identical to OK in this problem (see the left panels). Therefore, we need to choose $\tilde{\bm c}$ different from the Perron eigenvector to capture the heteroskedasticity. This is discussed in the next section.

\section{Heteroskedastic rational kriging}
\label{sec:hrk}
Since $\tau(\bm x)=\nu/(c_0+\bm r(\bm x)'\bm c)$, we can choose $\tilde{\bm c}=(c_0,\bm c')'$ appropriately to model the nonconstant variance of the response surface. This can be done by treating $\tilde{\bm c}$ as hyperparameters in (\ref{eq:RKmodel}) and jointly estimating them along with the other hyperparameters $\mu$, $\nu$, and $\bm \theta$. To obtain their estimates, we will use empirical Bayes methods, which is common in the analysis of computer experiments \citep{santner2019design}. 

The marginal distribution of the data is given by
\begin{equation}
    \Y_n|\Tilde{\bm{c}},\nu^2, \mu, \bm{\theta} \sim \mathcal{N}\left(\mu\bm{1}_n, \nu^2\{\text{diag}(\bm{d})\}^{-1}\R\{\text{diag}(\bm{d})\}^{-1}\right),
\end{equation}
where $\bm{d}=(d_1,\dots,d_n)'=\Tilde{\R}\Tilde{\bm{c}}=c_0\bm 1_n+\bm R\bm c$ and $diag(\bm d)$ is a diagonal matrix with diagonal elements $\bm d$. Thus, the maximum likelihood estimates (MLE) of the unknown hyperparameters can be obtained by maximizing
\begin{align*}
L &= p(\Y_n|\bm{X}_n,\Tilde{\bm{c}},\nu^2, \mu, \bm{\theta}) \nonumber \\
&\propto \frac{\prod_{i=1}^nd_i}{\nu^n|\bm R|^{1/2}}\exp\left\{-\frac{1}{2\nu^2}(\Y_n-\mu\bm{1}_n)'\text{diag}(\bm{d})\R^{-1}\text{diag}(\bm{d})(\Y_n-\mu\bm{1}_n)\right\}\nonumber
\end{align*}
or equivalently maximizing
\begin{align}
    \log L=&-\frac{n}{2}\log{\nu^2}+ \sum_{i=1}^n \log{d_i} - \frac{1}{2}\log{|\R|} \nonumber\\
    &-\frac{1}{2\nu^2}(\Y_n-\mu\bm{1}_n)'\text{diag}(\bm{d})\R^{-1}\text{diag}(\bm{d})(\Y_n-\mu\bm{1}_n).\label{eq:logL}
\end{align} 
Maximizing $\log L$ with respect to  $\mu$ and $\nu^2$, we obtain
\begin{align}
    \hat{\mu} &= \frac{\bm{d}'\bm{R}^{-1}\text{diag}(\bm{d})\Y_n}{\bm{d}'\bm{R}^{-1}\bm{d}},\label{eq:mu} \\ 
    \hat{\nu^2} &= \frac{1}{n}(\Y_n-\hat{\mu}\bm{1}_n)'\text{diag}(\bm{d})\R^{-1}\text{diag}(\bm{d})(\Y_n-\hat{\mu}\bm{1}_n).\label{eq:nu2}
\end{align}
Plugging them in (\ref{eq:logL}) gives 
\begin{equation}\label{eq:obj}
    \log L=-\frac{n}{2}\log{\hat{\nu}^2}+ \sum_{i=1}^n \log{d_i} - \frac{1}{2}\log{|\R|}-\frac{n}{2},
\end{equation}
which can be maximized with respect $\bm \theta$ and $\Tilde{\bm{c}}$.

To obtain a fast solution, we will first fit a rational kriging to the data and obtain $\bm \theta_{RK}$. Then we will numerically maximize $\log L$ with respect to  $\tilde{\bm c}$ with  $\tilde{\bm c}_{RK}$ as the initial value. This strategy simplifies the optimization task and enforces some sort of regularization to obtain a stable solution. Thus, the objective is to
\[\min_{\tilde{\bm c}} \log{\hat{\nu}^2}- \frac{2}{n}\sum_{i=1}^n \log{d_i}\]
subject to the constraints $\tilde{\bm c}'\tilde{\bm c}=1$ and $\tilde{\bm c}\ge 0$. This is a high-dimensional optimization with $n+1$ parameters and should therefore be handled carefully to obtain a computationally efficient solution.

First note that $\hat{\nu}^2$ can also be written as
\begin{eqnarray*}
    \hat{\nu}^2&=&\frac{1}{n}\bm d'\text{diag}(\Y_n-\hat{\mu}\bm{1}_n)\R^{-1}\text{diag}(\Y_n-\hat{\mu}\bm{1}_n)\bm d\\
    &=&\frac{1}{n}\tilde{\bm c}'\tilde{\bm R}' \text{diag}(\Y_n-\hat{\mu}\bm{1}_n)\R^{-1}\text{diag}(\Y_n-\hat{\mu}\bm{1}_n)\tilde{\bm R}\tilde{\bm c}
\end{eqnarray*}
which is a quadratic function in $\tilde{\bm c}$ if we fix $\hat{\mu}$ to a reasonable value. Thus, by fixing $\hat{\mu}=\mu_{RK}$, we obtain
\[\hat{\nu}^2=\frac{1}{n}\left\{ c_0^2\bm 1_n'\bm Q\bm 1_n+2c_0\bm c'\bm R\bm Q\bm 1_n+\bm c'\bm R\bm Q\bm R\bm c\right\},\]
where $\bm Q=\text{diag}(\Y_n-\mu_{RK}\bm{1}_n)\R^{-1}\text{diag}(\Y_n-\mu_{RK}\bm{1}_n)$. Since $c_0=\sqrt{1-\bm c'\bm c}$, the optimization problem becomes
\begin{align}
   \min_{\bm c} g(\bm c) =& \log{\left(\frac{1}{n}\left\{ (1-\bm c'\bm c)\bm 1_n'\bm Q\bm 1_n+2\sqrt{1-\bm c'\bm c}\;\bm c'\bm R\bm Q\bm 1_n+\bm c'\bm R\bm Q\bm R\bm c\right\}\right)} \nonumber\\
    &-\frac{2}{n}\bm 1_n'\log{\left(\sqrt{1-\bm c'\bm c} \;\bm 1_n+\bm R\bm c \right)}\label{eq:simplified_obj}\\
    \textrm{subject to} & \;\;\bm c'\bm c\le 1 \;\textrm{and}\; \bm c\ge 0, \nonumber
\end{align}
where $\log \bm x$ denotes $[\log x_1,\ldots,\log x_n]'$. The gradient of $g(\bm c)$ can be explicitly obtained as
\begin{align*}
    \frac{\partial g}{\partial \bm c} &= \frac{2}{n\hat{\nu}^2}\left\{ \bm R\bm Q\bm R\bm c+\sqrt{1-\bm c'\bm c} \;\bm R\bm Q\bm 1_n-\bm 1_n'\bm Q\bm 1_n\bm c-\frac{\bm c'\bm R\bm Q\bm 1_n}{\sqrt{1-\bm c'\bm c}}\bm c\right\}\\
    &-\frac{2}{n}\left( \bm R-\frac{\bm c\bm 1_n'}{\sqrt{1-\bm c'\bm c}}\right)\left\{ diag(\bm d)\right\}^{-1}\bm 1_n,
\end{align*}
where $\bm d= \sqrt{1-\bm c'\bm c} \;\bm 1_n+\bm R\bm c$. Now optimization can be carried out efficiently using the globally-convergent method-of-moving-asymptotes (MMA) algorithm \citep{svanberg2002class} through the Nlopt package \citep{nloptr} subject to constraints $\bm c'\bm c\le 1$ and $\bm c\ge 0$. The complete algorithm is summarized in Algorithm \ref{alg:optim}. It is important to note that $\bm R$ and $\bm R^{-1}$ are computed only once outside the optimization loop, making this algorithm extremely fast compared to the other nonstationary GPs.

\begin{algorithm}
   \caption{HRK Parameter Estimation Procedure}
   \label{alg:optim}
   \begingroup
   \setstretch{1.5}
   \begin{algorithmic}[1]
   \Require Design points $\bm{X}_n$, response $\Y_n$ 
   \setlength{\itemindent}{2em}
   \State Fit rational kriging to get $\bm{\theta}_\text{RK}, \mu_\text{RK}, \tilde{\bm c}_{RK}$;
   \State Minimize $g(\bm c)$ in \eqref{eq:simplified_obj} by MMA algorithm with $\bm c_{RK}$ as the initial value to get $\Tilde{\bm{c}}$ ;
   \State  Calculate $\hat{\mu}$ and $\hat{\nu}^2$ using \eqref{eq:mu} and \eqref{eq:nu2} respectively;
   \State \textbf{Return:} $\bm{\theta}, \hat{\mu}, \Tilde{\bm{c}}, \hat{\nu}^2$.
\end{algorithmic}
\endgroup
\end{algorithm}

Our approach is closely related to the maximum likelihood estimation proposed in \cite{kang2016kernel} for an iterated kernel regression. The important differences from their work are that (i) HRK contains $c_0$ making uncertainty quantification more stable, as shown in \cite{joseph2024rational} and (ii) we use a full gradient-based optimization which is extremely fast compared to the iterative nonlinear optimization used in \cite{kang2016kernel}.

Consider again the example of an underdamped oscillation system. The prediction and confidence intervals for heteroskedastic rational kriging (HRK) are shown in the top right panel of Figure \ref{fig:ExpTrig_fit}. We can see that the width of the confidence interval provided by HRK is decreasing to the right. The relative variance $\tau(\bm{x})$ shown in the bottom right panel exhibits a decreasing trend from $0$ to $1$, which agrees with our intuition. 

\begin{figure}
    \centering
    \includegraphics[width=0.8\textwidth]{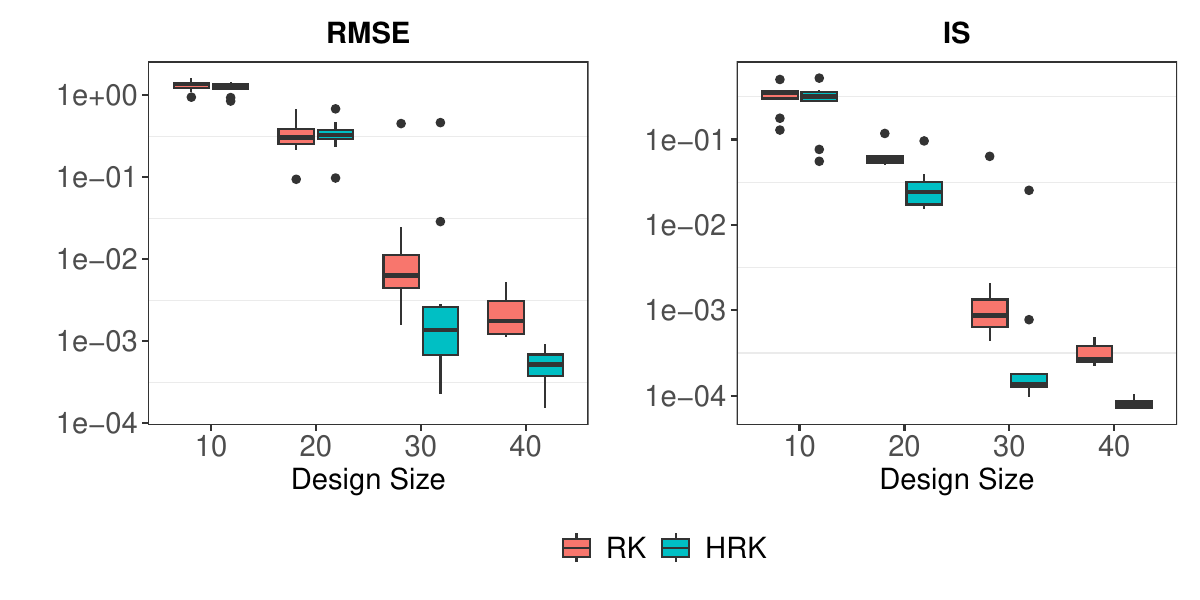}
    \caption{Comparison of RK and HRK as surrogate model for the underdamped oscillation system. Boxplots of RMSE and IS are generated from 10 random Latin hypercube designs of size $10,\,20,\,30,\,40$.}
    \label{fig:ExpTrig_metric}
\end{figure}

To quantitatively compare the performance of different methods, we use two metrics: root mean-squared error (RMSE) to evaluate prediction performance and interval score (IS) \citep{gneiting2007strictly} to assess uncertainty quantification. The interval score is defined as
\begin{equation*}
    IS = \frac{1}{N} \sum_{i=1}^N\left[(u_i-l_i)+\frac{2}{\alpha}\left\{(l_i-t_i)_{+}+(t_i-u_i)_{+}\right\}\right],
\end{equation*}
where $(x)_{+} = x$ if $x>0$ and $0$ otherwise, $[l_i,\,u_i]$ is the $(1-\alpha)$ confidence interval, and $t_i$ is the true function values at the $i$th testing location, $i=1,\ldots,N$.
We construct a random Latin hypercube design (LHD) of sizes 10 to 40 and fit RK and HRK. For each design size, we repeat the experiment 10 times. The results are shown as boxplots in Figure \ref{fig:ExpTrig_metric}. It demonstrates that HRK consistently outperforms RK, offering noticeable improvements in both prediction and uncertainty quantification.  

\section{Active learning}\label{sec:active learning}
In this section, we will describe the application of heteroskedastic rational kriging (HRK) in active learning. Consider the ALM strategy discussed in the introduction: 
\begin{equation}\label{eq:ALM-HRK}
    \x_{n+1} = \argmax_{\x\in[0,1]^p} \frac{{1-\bm{r}(\x)'\R^{-1}\bm{r}(\x)}}{\{c_0+\bm{r}(\x)'\bm{c}\}^2}.
\end{equation}
The numerator is the same as the variance of OK, which will ensure that the points have a good space-filling property in the space of important variables. The denominator captures the inverse of the variance, which will ensure that the points are placed in regions that are difficult to predict. Thus, HRK enforces these dual objectives in the ALM strategy.  

Finding the next design point using (\ref{eq:ALM-HRK}) involves continuous optimization in $p$ dimensions, which can be time consuming. To simplify optimization, we search for the next design point on a candidate set of size $100(p+1)^2$. This strategy works well as long as we have a good candidate set. For this purpose, we first generate a random
Latin hypercube design of size $1000(p+1)^2$ and down-sample it to $100(p+1)^2$ points using the twinning algorithm \citep{vakayil2022data}, which is a quick way to construct a large space-filling design. There are better candidate generation strategies \citep{wycoff2024voronoi}, but they can be slower than the foregoing strategy. The complete algorithm is described in Algorithm \ref{alg:ActiveLearning}.

\begin{algorithm}
   \caption{Active Learning with HRK}
   \label{alg:ActiveLearning}
   \begingroup
   \setstretch{1.5}
   \begin{algorithmic}[1]
   \Require Experiment budget $n$, initial design size $n_{\text{ini}}$, computer model $f$.
   \State Construct initial design $\bm{X}_{n_{\text{ini}}}$
   \State Evaluate the the responses $\Y_{n_{\text{ini}}} \leftarrow [f(\x_1),\ldots,f(\x_{n_{\text{ini}}})]$
   \State $\bm{D} \leftarrow [(\x_1, y_1), \dots, (\x_{n_{\text{ini}}}, y_{n_{\text{ini}}})]$
   \For{$i$ in $(n_{\text{ini}}+1),\dots,n$}
       \State Fit HRK on $\bm{D}$ by Algorithm \ref{alg:optim} 
       \State Generate candidate points $\bm{X_\text{cand}}$
       \State $\x_{i}=\argmax_{\x\in\bm{X_\text{cand}}}s^2(\x)$
       \State Augment the dataset $\bm{D} \leftarrow \bm{D}\cup \{(\x_{i}, y_{i})\}$
   \EndFor
   \State \textbf{Return:} $\bm{D}$
\end{algorithmic}
\endgroup
\end{algorithm}

\begin{figure}
    \centering
    \includegraphics[width=0.75\textwidth]{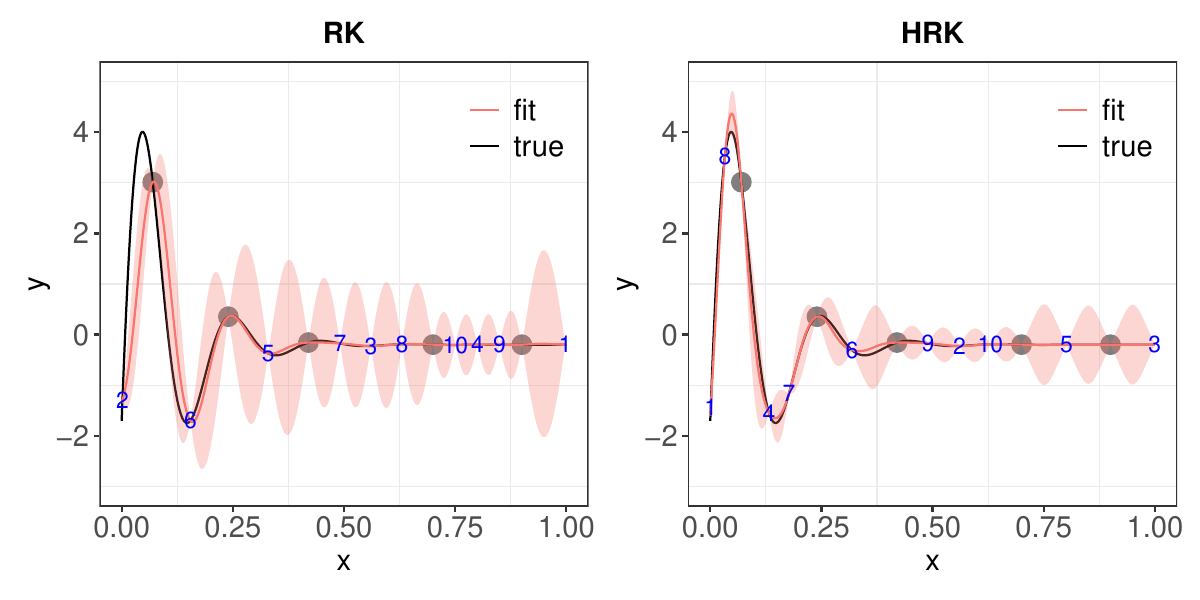}
    \caption{Active learning based on RK and HRK on the underdamped oscillation system. Gray points are the initial 5 design points and blue numbers denotes the next 10 design points selected by ALM. The red solid lines denote the predicted mean and the red shaded bands are the 95\% confidence intervals. Black line is the true response.}
    \label{fig:ExpTrig_seq}
\end{figure}

Consider again the underdamped oscillation system discussed in the Introduction.
The right panel of Figure \ref{fig:ExpTrig_seq} shows 10 points (blue numbers) selected using Algorithm \ref{alg:ActiveLearning} starting with an initial design of five points (gray points). The same is done using RK and is shown in the left panel of the figure. We can see that the ALM startegy using HRK allocates more points to the left side of the function than that of RK. For example, out of the 10 points, four points are allocated to $x<0.25$ with HRK, whereas only two points are allocated to the same region with RK. As a result, HRK is able to approximate the function in a better way than RK.

\begin{figure}[!h]
    \centering
    \begin{subfigure}[c]{0.65\textwidth}  
        \centering 
        \includegraphics[width=\textwidth]{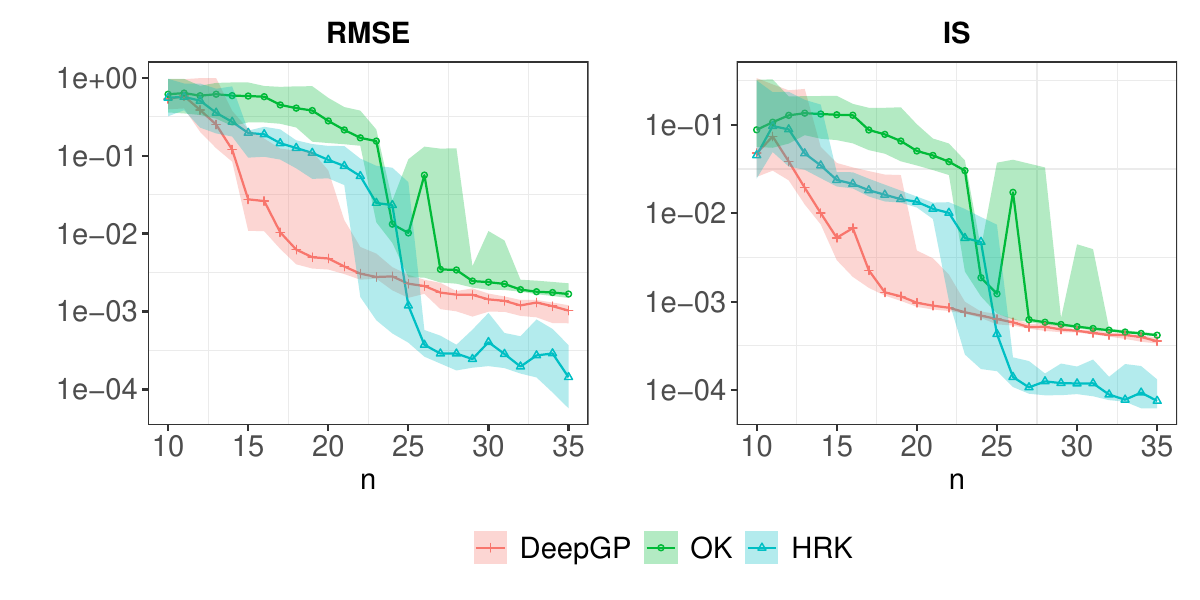}
        \caption{}    
        \label{fig:ExpTrig_seqD}
    \end{subfigure}
    \begin{subfigure}[c]{0.31\textwidth}  
        \centering 
        \includegraphics[width=\textwidth]{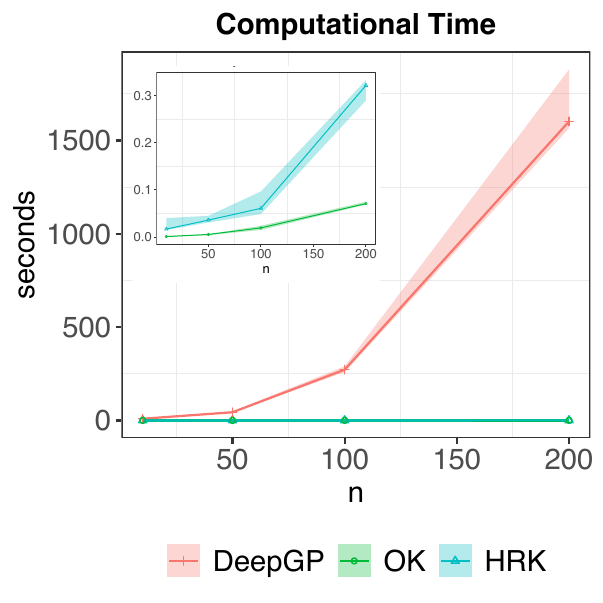}
        \caption{}    
        \label{fig:compute_time}
    \end{subfigure}
    \caption{(a) Active learning based on OK, HRK, and DeepGP. The initial designs are random LHDs of size 10. Solid lines represent the median over 10 repetitions and the shaded bands mark the 5th and 95th quantiles. (b) Computational time for fitting DeepGP, HRK, and OK on underdamped oscillation system with $n$ varied from $10$ to $200$. A zoomed-in view is shown as an inset to see the difference between OK and HRK.}
\end{figure}

To evaluate HRK's performance on active learning for the underdamped oscillation example, we compare it against two methods: ordinary kriging using R package: \texttt{rkriging}  \citep{rkrigingRpackage} and deep Gaussian process using R package: \texttt{deepgp}  \citep{deepgpRpackage}. For DeepGP, we adopt a two-layer structure and follow the algorithmic specifications in \cite{sauer2023active}. For all three methods, we initialize the active learning process using a random Latin hypercube design of 10 points and sequentially add 25 points. The results are shown in Figure \ref{fig:ExpTrig_seqD}. We can see that in this example HRK performs better than OK in terms of prediction and uncertainty quantification and also surpasses DeepGP in the end. 

A major advantage of HRK is its fast computation compared to other nonstationary GPs. The computational time of HRK, deep Gaussian process, and OK for fixed designs of different sizes is shown in Figure \ref{fig:compute_time}, where the simulation is performed on a $2.6$ GHz desktop with $16.0$ GB memory. We can see that HRK is more than three orders of magnitude faster than DeepGP and is only slightly slower than OK.

\section{Simulations}\label{sec:simulation}
\begin{figure}
    \centering
    \includegraphics[width=0.8\textwidth]{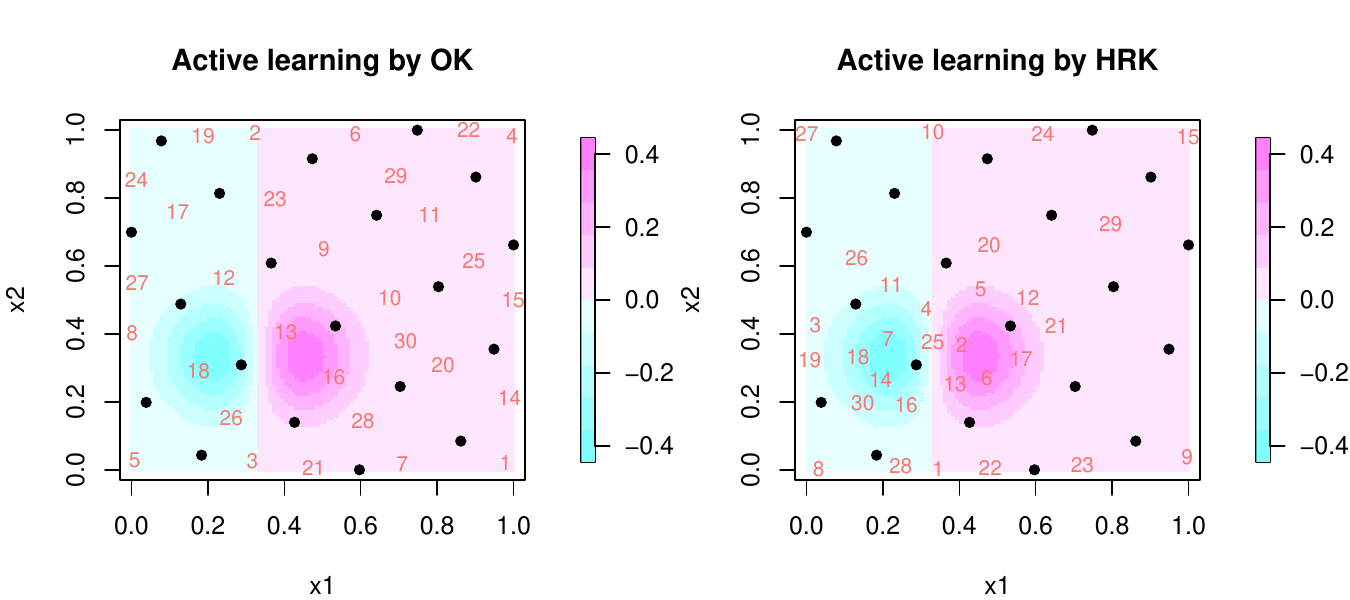}
    \caption{Design points generated using active learning based on ordinary kriging and heteroskedastic rational kriging for the Gramacy-Lee function, where the input domain is scaled to $[0, 1]^2$. The 20 initial points by a MaxPro design are shown as black dots, while the 30 sequentially designed points are labeled as numbers.}
    \label{fig:gramacy_seqD_visual}
\end{figure}

We now investigate the numerical performance of HRK against OK and DeepGP on several test functions. Consider the two-dimensional function from \cite{gramacy2009adaptive}:
\begin{equation}\label{eq:gramacy}
    f (x_1, x_2) = x_1\exp{(-x_1^2-x_2^2)} \quad \text{for} \;x_1, x_2 \in [-2, 4].
\end{equation}
This function exhibits a distinct variation pattern in the domain: it changes rapidly near the lower left corner, whereas the rest of the domain remains relatively flat. Starting with a MaxPro design \citep{joseph2015maximum} of 20 points, we add 30 points sequentially using ALM strategy. Figure \ref{fig:gramacy_seqD_visual} shows the points obtained using OK (left panel) and HRK (right panel).  We can see that the ALM strategy using OK produces points that look almost like a maximin-type space-filling design. On the other hand, HRK allocates more points near the region of high variability, which can lead to better approximation of the response surface.

Now consider some higher-dimensional functions: 8-dimensional Dette-Pepelyshev function \citep{dette2010generalized}, 4-dimensional cantilever beam function \citep{eldred2007investigation}, 6-dimensional OTL circuit function \citep{ben2007modeling}, 7-dimensional piston function \citep{kenett1998modern}, and 8-dimensional borehole function \citep{morris1993bayesian}. Details of these test functions are available at the Virtual Library of Simulation Experiments website \citep{simulationlib}. 

\begin{figure}
    \begin{subfigure}[t]{0.49\linewidth}
        \includegraphics[width=\linewidth]
        {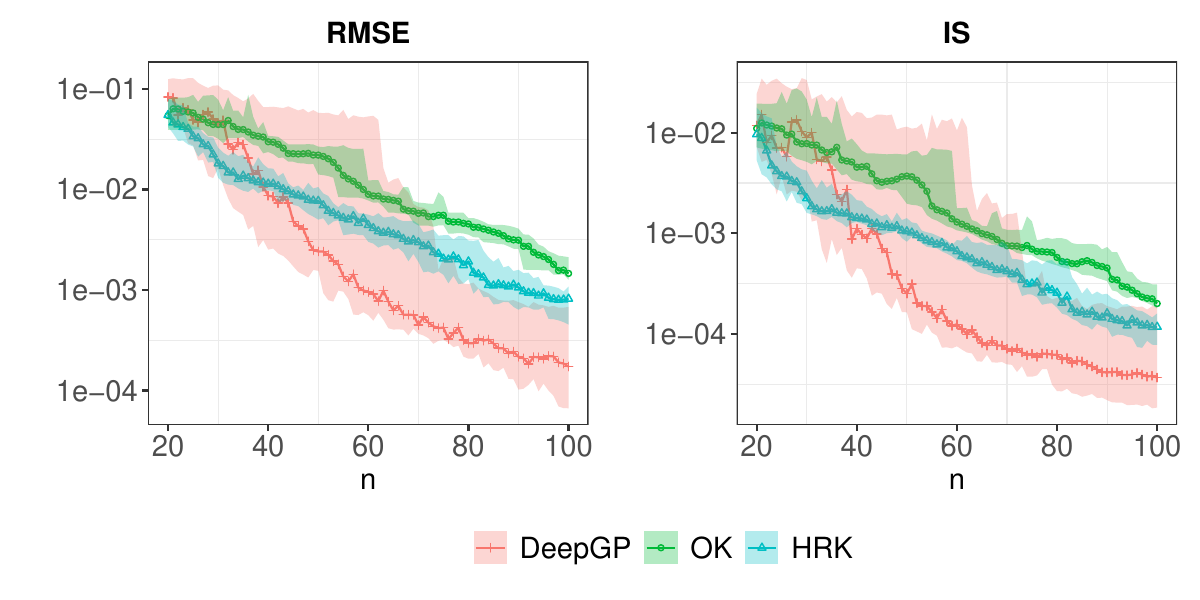}
        \caption{Gramacy-Lee function}
        \label{fig:gramacy_seq}
    \end{subfigure}
    \begin{subfigure}[t]{0.49\linewidth}
        \includegraphics[width=\linewidth]
        {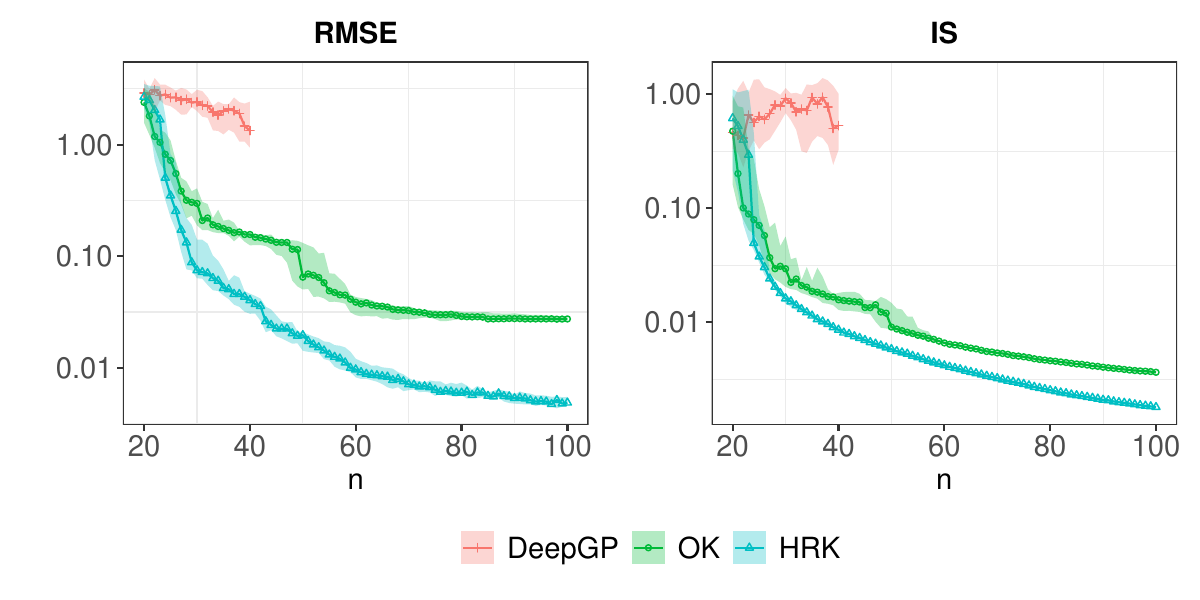}
        \caption{Dette-Pepelyshev function}
        \label{fig:detpep10curv_seq}
    \end{subfigure}
    \begin{subfigure}[t]{0.49\linewidth}
        \includegraphics[width=\linewidth]{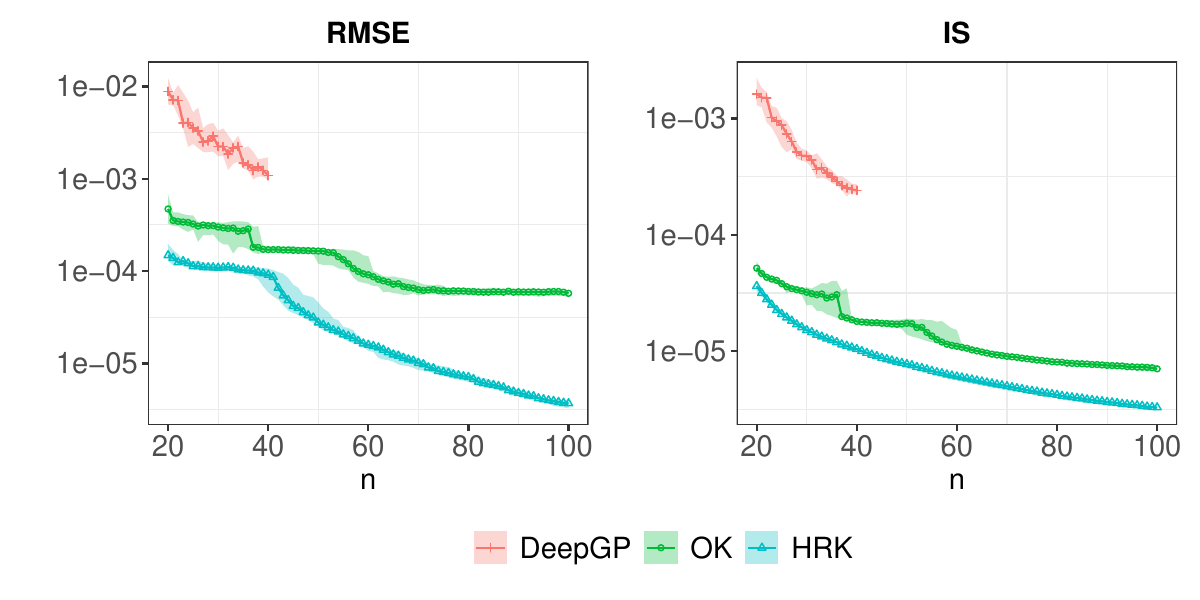}
        \caption{Cantilever beam function}
        \label{fig:canti_seq}
    \end{subfigure}
    \begin{subfigure}[t]{0.49\linewidth}
        \includegraphics[width=\linewidth]{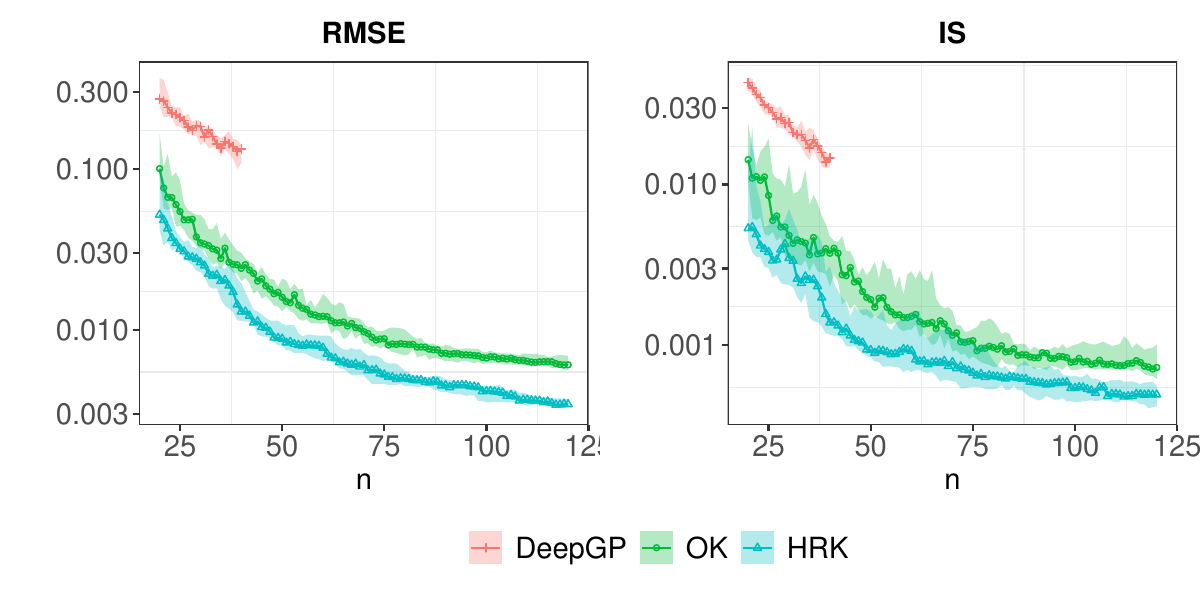}
        \caption{OTL circuit function}
        \label{fig:olt_seq}
    \end{subfigure}
        \begin{subfigure}[t]{0.49\linewidth}
        \includegraphics[width=\linewidth]{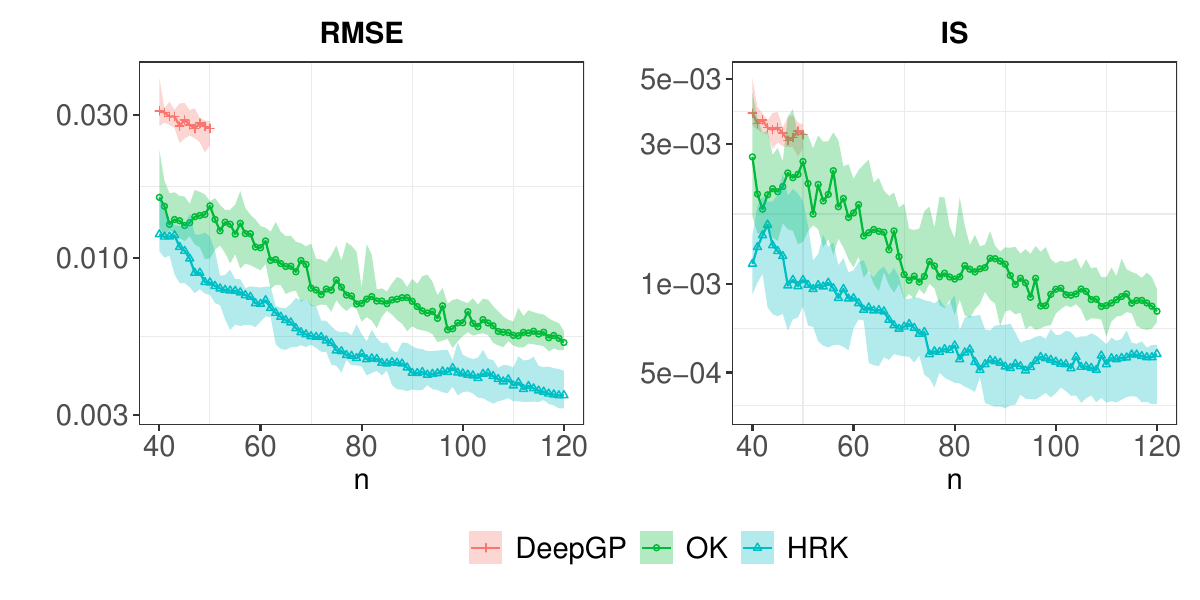}
        \caption{Piston function}
        \label{fig:piston_seq}
    \end{subfigure}
    \begin{subfigure}[t]{0.49\linewidth}
        \includegraphics[width=\linewidth]{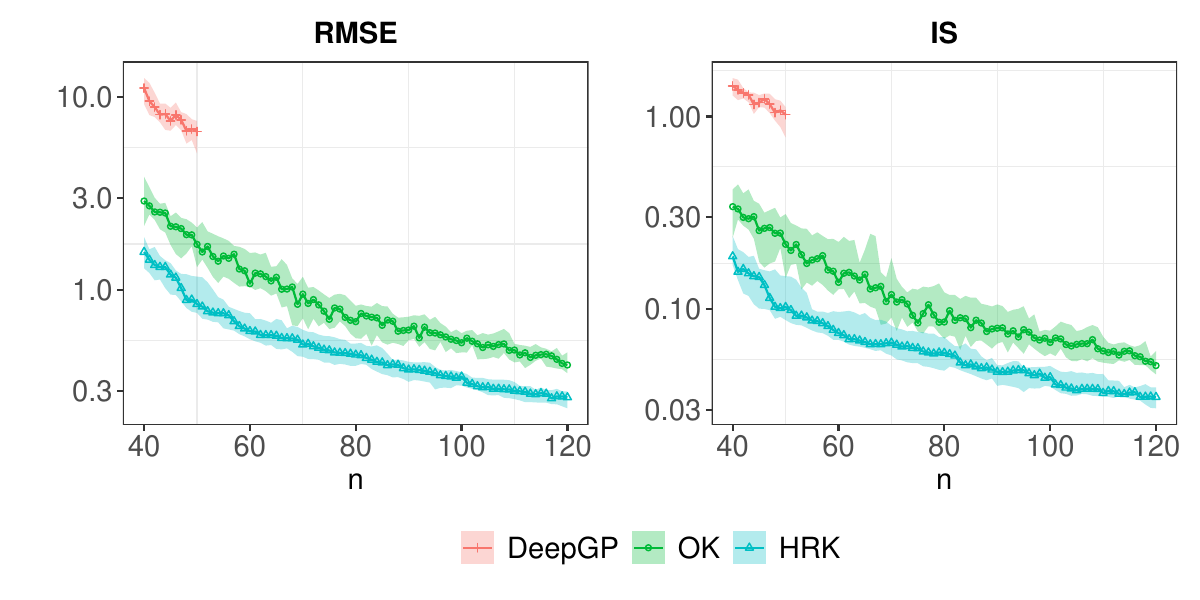}
        \caption{Borehole function}
        \label{fig:borehole_seq}
    \end{subfigure}
    \caption{Performance comparison for active learning with HRK, DeepGP and OK. The initial designs are random LHD of size $10p$. Solid lines represent the median over 10 repetitions and the shaded bands mark the 5th and 95th quantiles.}
    \label{fig:result_seqD}
\end{figure}

 The results of the simulations with the test functions described above are summarized in Figure \ref{fig:result_seqD}. DeepGP is run only for a few steps except for the 2-dimensional Gramacy-Lee function due to its high computational time and poor performance.  For almost all cases, HRK exhibits improved accuracy over OK. Although DeepGP outperforms HRK in the Gramacy-Lee function, this advantage is not consistent across all scenarios. On the other hand, HRK's performance seems to be more stable than that of DeepGP.  Furthermore, HRK’s simpler structure makes it more interpretable and computationally efficient.

\section{Real Datasets}\label{sec:case}
In this section, we compare the active learning performances of HRK against OK and DeepGP using two real datasets.

\subsection{Sublimation Enthalpy Prediction}
Sublimation enthalpy ($\Delta H_\text{sub}$) is a key thermodynamic parameter that quantifies the energy required to sublimate a material. It plays a critical role in hybrid membrane process techniques \citep{leng2017vapor}, as it determines the temperature needed to volatilize an organic precursor to achieve a partial pressure suitable for the  vapor phase infiltration process. Density Functional Theory (DFT) \citep{hohenberg1964inhomogeneous} is the state-of-the-art method for calculating sublimation enthalpy with high accuracy. However, its significant computational cost makes it impractical to evaluate every molecule of interest \citep{lee2010higher}. The development of comprehensive molecular descriptors/fingerprints for organic compounds and polymers has enabled the use of statistical models trained on DFT data to predict material properties efficiently. A better selection of molecule structures to perform DFT is therefore important for the statistical model to quickly achieve high accuracy. 

\begin{figure}
    \centering
    \includegraphics[width=0.75\linewidth]{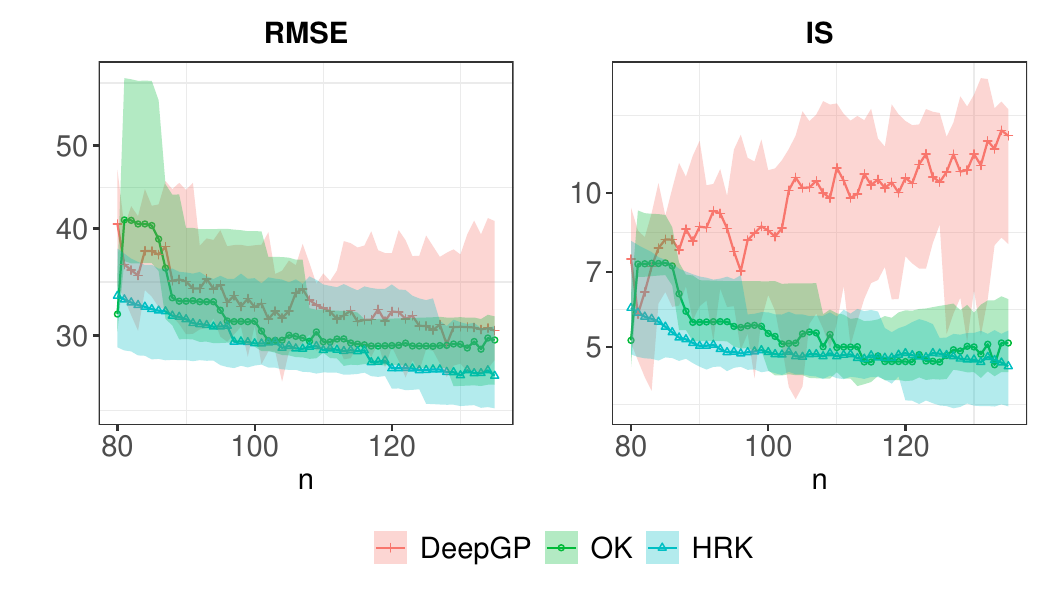}
    \caption{RMSE and IS trajectories for sublimation enthalpy prediction as new molecule structures are selected by active learning.}
    \label{fig:Enthalpy_seqD}
\end{figure}

\cite{liu2024} generated a dataset comprised of 815 unique organic molecules of interest: https://khazana.gatech.edu/dataset/.  In order to demonstrate the utility of active learning methods, we will use this dataset to sequentially select structures to run the DFT model for the sublimation enthalpy calculation. The molecular structures are first transformed by a cheminformatics software (RDkit) into descriptors \citep{rdkit}. Five principal components are calculated from these descriptors as input for the following three models: ordinary kriging, deep Gaussian process, and heteroskedastic rational kriging. We start with 80 structures and then sequentially add 55 new structures from the dataset through active learning. The unselected structures are used as test points to evaluate the models' accuracy. The results are shown in Figure \ref{fig:Enthalpy_seqD}, where HRK shows consistent improvement over both OK and deep Gaussian process.

\subsection{Langley Glide-Back Booster Dynamics}
Langley Glide-Back Booster (LGBB) Dynamics was a project proposed by the National Aeronautics and Space
Administration (NASA) to develop a reusable rocket booster \citep{rogers2003automated, gramacy2009adaptive}. They aimed to study the flight characteristics (lift, drag, pitch, side-force, yaw, and roll) of the LGBB during atmospheric reentry, with a focus on three key inputs: speed (Mach number), angle of attack (alpha), and side-slip angle (beta). Computational fluid dynamics (CFD) simulations were conducted to explore the rich dynamics under different design configurations. 

\begin{figure}[h]
    \centering
    \begin{subfigure}[b]{0.35\linewidth}
        \includegraphics[width=\linewidth]{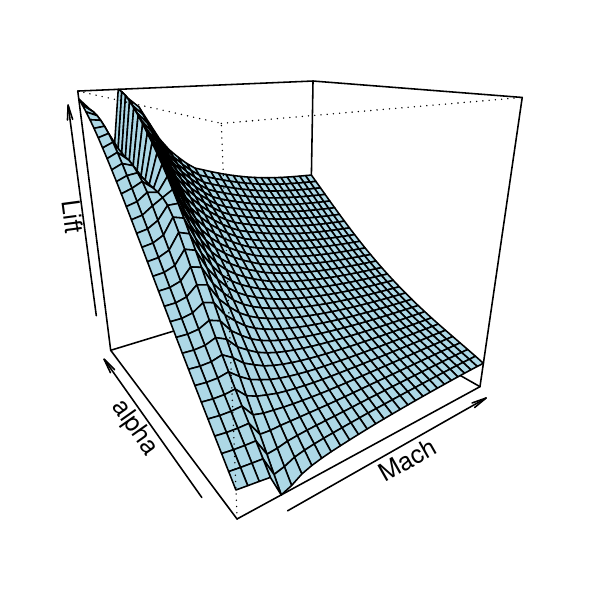}
        \caption{}
        \label{fig:lgbb_surf}
    \end{subfigure}
    \begin{subfigure}[b]{0.64\linewidth}
        \includegraphics[width=\linewidth]{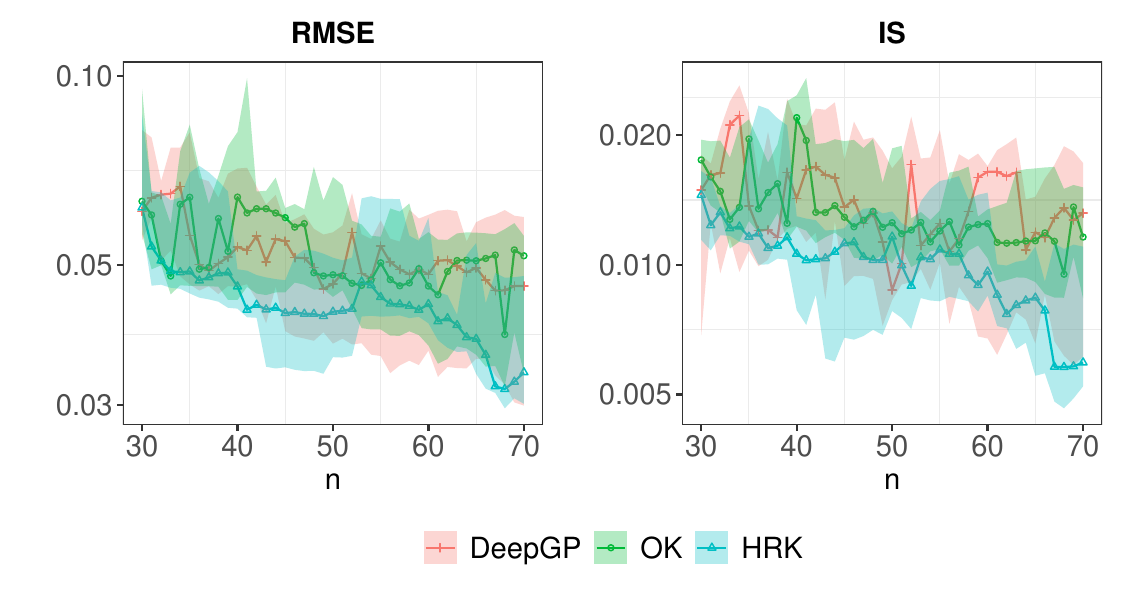}
        \caption{}
        \label{fig:lgbb_lift_seqD}
    \end{subfigure}
    \caption{(a) Response surface of the lift as a function of Mach and angle of attack (alpha) with side-slip
angle (beta) fixed as 1; (b) RMSE and IS trajectories as new design points are added by active learning.}
\end{figure}

The dataset is obtained from the lgbb archive at https://bobby.gramacy.com/surrogates/. We interpolate the 37,908 data points using Treed Gaussian process \citep{gramacy2008bayesian} and use it as the ground truth. Here we focus only on the lift output, as the other outputs exhibit similar dynamics. We sample 3,000 Sobol points as test locations. All models are trained on a common initial random Latin hypercube design of size 30, and then 40 additional points are selected by Algorithm \ref{alg:ActiveLearning}. The results of ten replicated experiments are presented in Figure \ref{fig:lgbb_lift_seqD}, where HRK demonstrates a clear improvement over OK in terms of both the RMSE and IS metrics. This could be because, as shown in Figure \ref{fig:lgbb_surf}, lift has a complex transition behavior near the subsonic and supersonic boundary and a small variation throughout the  supersonic region. Interestingly, DeepGP did not exhibit a performance advantage over OK in this scenario, potentially due to suboptimal Bayesian parameter inference. 


\section{Conclusion}\label{sec:remarks}
For emulating complex computer models, the main task is to gather data at the right locations in the experimental region so that we can get a good approximation of the response surface. The optimal locations to run the computer codes (optimal experimental design) can be determined based on the potential prediction errors of the emulator (or surrogate model). GP/kriging provides a nice framework for doing this because of its ability to provide uncertainty quantification of prediction errors, which are measured through the posterior variance of the underlying function. Active learning procedures such as ALM work by placing the design points sequentially at locations where the posterior variance is a maximum. Unfortunately, stationarity assumption of variance is commonly employed in GP/kriging, which makes the ALM points behave almost like a maximin-type design, which tends to uniformly fill the entire experimental region. This does not constitute an efficient use of resources because some regions of the response surface can be harder to predict and demand a larger allocation of design points. In this article, we attempted to overcome this deficiency by using a heteroskedastic rational kriging model that captures the nonconstant variance of the response surface. Simulations with various test functions show that HRK is able to produce comparable or better performance in ALM while being orders of magnitude faster than other nonstationary-based emulation techniques.

This article focuses on active learning procedures for emulation. An important future direction for research is to explore active learning procedures when the objective is optimization. Since HRK is able to provide improved uncertainty quantification, we expect that it will do well in providing a better exploration/exploitation trade-off in the active learning procedures aimed at optimization, but this needs further investigation.

\vspace{.25in}
\noindent {\Large\bf Acknowledgments}

\noindent This work is supported by U.S. National Science Foundation grants DMREF-1921873 and DMS-2310637.

\bibliographystyle{asa}
\bibliography{Bibliography.bib}

\end{document}